\documentclass[particles,article,accept,pdftex,moreauthors]{Definitions/mdpi}

\firstpage{1}
\pubvolume{1}
\issuenum{1}
\articlenumber{0}
\pubyear{2023}
\copyrightyear{2023}
\externaleditor{Academic Editor: Evgeni Kolomeitsev, Nikolai Antonenko, David Blaschke, Victor Braguta, Isaac Vidaña}
\datereceived{02 June 2023 }
\daterevised{ } 
\dateaccepted{25 June 2023 }
\datepublished{ }
\hreflink{https://doi.org/} 

\Title{Neutrino Spectrum and Energy Loss Rates Due to Weak Processes on Hot $^{56}$Fe in  Pre-Supernova Environment}

\TitleCitation{Neutrino Spectrum and Energy Loss Rates Due to Weak Processes on Hot $^{56}$Fe in  Pre-Supernova Environment}

\Author{A. A. Dzhioev $^{1,}$*\orcidA{}, A. V. Yudin $^{2}$\orcidB{}, N. V. Dunina-Barkovskaya $^{2}$  and A. I. Vdovin  $^{1}$}

\AuthorNames{Alan A. Dzhioev, A. V. Yudin,  N. V. Dunina-Barkovskaya and  A. I. Vdovin}

\AuthorCitation{Dzhioev, A.A.; Yudin, A.V.; Dunina-Barkovskaya, N.V.; Vdovin, A.I.}

\address{%
$^{1}$ \quad Bogoliubov Laboratory of Theoretical Physics, Joint Institute for Nuclear Research, 141980 Dubna, Russia; vdovin@theor.jinr.ru\\
$^{2}$ \quad National Research Center “Kurchatov Institute”, 123182 Moscow, Russia; yudin@itep.ru (A.V.Y.); dunina@itep.ru (N.V.D.-B.)\\
}

\corres{Correspondence: dzhioev@theor.jinr.ru; Tel.: +7-49621-63-385}

\abstract{Applying TQRPA calculations of Gamow--Teller strength functions in hot nuclei, we compute the (anti)neutrino spectra and  energy loss rates arising from weak processes on hot $^{56}$Fe under pre-supernova conditions. We use a realistic pre-supernova model calculated by the stellar evolution code MESA.  Taking into account both charged and neutral current processes, we demonstrate that weak reactions with hot nuclei can  produce high-energy (anti)neutrinos. We also show that, for hot nuclei, the energy loss via (anti)neutrino emission is significantly larger than that for nuclei in their ground state.   It is found that the neutral current de-excitation via  the $\nu\bar\nu$-pair emission is presumably a dominant source of antineutrinos. In accordance with other studies,   we confirm that the so-called single-state approximation for neutrino spectra might fail under certain pre-supernova conditions.    }


\begin{document}


\section{Introduction}

It is well known that the production and propagation of (anti)neutrinos in the stellar matter are important ingredients of the computer modeling of stellar evolution. According to the theory, in~stellar interiors with both high temperature and density, neutrino  emission makes a major contribution to energy loss,  removes entropy from the stellar core and accelerates  the evolution of the star~\cite{Woosley_RMPh74,Janka_PhysRep442,Balasi_PPNP85}. The~observation of neutrinos from supernova SN1987A confirmed and advanced our understanding  of core-collapse supernova explosion. Recent remarkable progress in neutrino detection techniques may enable     the registration of neutrinos from new sources. Some of the candidates  are pre-supernova (anti)neutrinos emitted from the core of a massive star just before the collapse~\cite{Kato_ARNPS70}. Although~pre-supernova   (anti)neutrinos have not been detected to date, their observation would offer a possibility for studying the physical processes that lead to core collapse and would be a warning of an upcoming~explosion.

In~\cite{Patton_ApJ840,Patton_ApJ851}, the~role of charged current nuclear weak processes (electron  and positron capture, $\beta^\mp$-decay) in the neutrino emission from a pre-supernova star was studied. It was found that, under certain conditions, nuclear processes compete with thermal processes (plasmon decay, pair annihilation, etc.) in their contribution to the (anti)neutrino flux or even dominate in the energy window relevant for detection. However, it was pointed out that, while total emissivities are relatively robust,  the~highest-energy tails of the neutrino spectrum, in~the detectable window, are very sensitive to the details of the calculations. Specifically, the~source of the error lies in the single-strength approximation~\cite{Langanke_PRC64} that was adopted in~\cite{Patton_ApJ840,Patton_ApJ851} for nuclear processes. In~\cite{Misch_PRC94}, an~exploratory study of this error was performed and it was shown that the specific neutrino spectrum obtained from the single-strength approximation could miss important~features.

High-temperature stellar plasma allows  nuclei to access excited states in accordance with the Boltzmann distribution.   In~\cite{Langanke_PRC64},  the~single-strength approximation was derived assuming that (i) weak processes on a thermally excited state in the parent nucleus lead to the Gamow--Teller (GT) transition to a single state in the daughter nucleus and that (ii) the Brink hypothesis is valid, i.e.,~the GT strength function is the same for all excited states.
The violation of the  Brink hypothesis for thermally excited (hot) nuclei was demonstrated for both charge-exchange~\cite{Dzhioev_PRC81,Dzhioev_PRC92,Dzhioev_PRC101} and charge-neutral~\cite{Dzhioev_PRC89,Dzhioev_PRC94} Gamow--Teller strength functions, and~it was shown that, under certain stellar conditions, thermal effects  on the GT strength significantly affect the rates and cross-sections of the nuclear weak process (as can also be seen in recent reviews~\cite{Dzhioev_PhPN53_1,Dzhioev_PhPN53_2,Dzhioev_PhPN53_3}).

In this paper, we apply the formalism of~\cite{Dzhioev_PRC81,Dzhioev_PRC92,Dzhioev_PRC101,Dzhioev_PRC89,Dzhioev_PRC94,Dzhioev_PhPN53_1,Dzhioev_PhPN53_2,Dzhioev_PhPN53_3} to study electron (anti)neutrino spectra and energy loss rates arising from weak processes on hot $^{56}$Fe under conditions realized in the pre-supernova environment. Besides~the charged current weak nuclear processes  considered in~\cite{Patton_ApJ840,Patton_ApJ851}, we also take into account  the neutral current de-excitation of  hot $^{56}$Fe via neutrino--antineutrino pair emission. The~main goal of the present work is to study how thermal effects on the GT strength function and $\nu\bar\nu$-pair emission  affects the (anti)neutrino spectra and energy loss~rates.

\section{Method}

To compute (anti)neutrino spectra and energy loss rates due to  weak processes  on hot nuclei, we apply a method which is based on the statistical
formulation of the nuclear many-body problem at finite temperature. In~this method, rather than compute GT strength distributions for individual thermally excited states, we determine a
thermal averaged strength function for the GT  operator
\begin{equation}\label{str_funct1}
  S_{\mathrm{GT}_{\pm, 0}}(E,T) = \sum_{i,f} p_i(T)B^{(\pm,0)}_{if}\delta(E-E_{if}),
\end{equation}
where $p_i(T)= e^{-E_i/kT}/Z(T)$ is the Boltzmann population factor for a parent state $i$  at a temperature $T$,  $B^{\pm,0}_{if} = |\langle f\|\mathrm{GT}_{\pm, 0}\|i\rangle|^2/(2J_i+1)$ is the reduced transition probability (transition strength) from the state $i$ to the state $f$ in the daughter nucleus; $\mathrm{GT}_0=\vec{\sigma}t_0$  for neutral current reactions and $\mathrm{GT}_\mp=\vec{\sigma}t_\pm$ for
charged current reactions. The~zero component of the isospin operator is denoted by $t_0$, while
$t_-$ and $t_+$ are the isospin-lowering ($t_-|n\rangle = |p\rangle$) and isospin-rising ($t_+|p\rangle = |n\rangle$) operators. Thus, `0' refers to the $\nu\bar\nu$-pair emission, `$-$' to positron capture (PC) and $\beta^-$-decay, and~`$+$' to electron capture (EC) and $\beta^+$-decay. The~transition energy between initial and final states  is given by $E_{if} = Q + E_f-E_i$, where $E_i$ and $E_f$ are the excitation energies of the parent and daughter nuclei, and~$Q=M_f-M_i$ is the ground-state reaction threshold (for neutral current reactions $Q=0$).   The definition of $S_\mathrm{GT}(E,T)$ implies that  at $T\ne 0$ the strength function is defined for both positive ($E>0$) and negative ($E<0$) energy domains.
The latter corresponds to the de-excitation of thermally excited states to states at lower energies. In~addition, low-energy transitions between excited states become possible at $T\ne 0$.

Obviously, the~explicit state-by-state calculation of  $S_{\mathrm{GT}_{\pm, 0}}(E,T)$ is hardly possible due to the extremely large  number
of nuclear states  thermally populated at stellar temperatures.  To~compute the temperature-dependent strength function \eqref{str_funct1},
we apply the TQRPA framework which is
a technique based on the quasiparticle random phase approximation (QRPA) extended to the
finite temperature by the superoperator formalism in the Liouville space~\cite{Dzhioev_PhPN53_1}. The~central concept of the TQRPA framework is the thermal vacuum
$|0(T)\rangle$, a~pure state in the  Liouville space, which corresponds to the grand canonical density matrix operator for the hot nucleus.
The time-translation operator in the Liouville space is the so-called thermal Hamiltonian $\mathcal{H}$ constructed from the nuclear Hamiltonian after introducing particle creation and annihilation superoperators. Within~the TQRPA,  the~strength function~\eqref{str_funct1} is
expressed in terms of the transition matrix elements from the thermal
vacuum to eigenstates (thermal phonons) of the thermal Hamiltonian $\mathcal H|Q_i\rangle= \omega_i|Q_i\rangle$:
\begin{equation}\label{str_funct2}
 S_{\mathrm{GT}_{\pm, 0}}(E,T) = \sum_i \mathcal{B}^{(\pm,0)}_{i}\delta(E-\omega_i \mp \Delta_{np}).
\end{equation}
Here, $\mathcal{B}^{(\pm,0)}_{i} = |\langle Q_i\|\mathrm{GT}_{\pm, 0}\|0(T)\rangle|^2$ is the transition strength to the $i$th state of a hot nucleus and $E_i^{(\pm, 0)} = \omega_i \pm \Delta_{np}$ is the transition energy; $\Delta_{np}=0$ for charge-neutral transitions, while  for charge-exchange transitions $\Delta_{np}=\delta\lambda_{np} + \delta M_{np}$, where $\delta\lambda_{np} = \lambda_n-\lambda_p$ is the difference between neutron and proton chemical potentials in the nucleus, and~$\delta M_{np}=1.293$\,MeV is the neutron--proton mass splitting. Note that  the eigenvalues of the thermal Hamiltonian,
$\omega_i$, take both positive and negative values. The~latter contribute to the strength function only at $T\ne 0$. We also stress that the strength function~\eqref{str_funct2} obeys the detailed balance principle:
\begin{equation}\label{DB1}
   S_{\mathrm{GT}_{0}}(-E,T)= \mathrm e^{-E/kT}  S_{\mathrm{GT}_{0}}(E,T)
\end{equation}
for charge-neutral GT transitions, and~\begin{equation}\label{DB2}
   S_{\mathrm{GT}_{\mp}}(-E,T)= \mathrm e^{-(E\mp\Delta_{np})/kT}  S_{\mathrm{GT}_{\pm}}(E,T)
\end{equation}
for charge-exchange GT transitions. This property makes the approach thermodynamically~consistent.

In what follows, we assume that emitted (anti)neutrinos freely leave the star. Then, we can write the following expressions for electron (anti)neutrino spectra resulting from the GT transition from the thermal vacuum to the $i$th state of a hot~nucleus:
\begin{itemize}
\item Electron or positron capture
\begin{multline}\label{spectrum_capture}
  \lambda_i^{\text{EC, PC}} (E_\nu)= \frac{G^2_\text{F} V^2_{\text{ud}} (g^*_A)^2}{2\pi^3\hbar^7c^6}\mathcal{B}^{(\pm)}_i (E_\nu +E^{(\pm)}_i)[(E_\nu +E^{(\pm)}_i)^2-m_e^2c^4]^{1/2}E^2_\nu
  \\
\times f_{e^{\mp}}(E_\nu +E^{(\pm)}_i) F(\pm Z, E_\nu +E^{(\pm)}_i)\Theta(E_\nu +E^{(\pm)}_i-m_ec^2),
\end{multline}
where upper (lower) sign corresponds to EC (PC);
\item $\beta^{\mp}$-decay
\begin{multline}\label{spectrum_decay}
  \lambda_i^{\beta^{\mp}} (E_\nu)= \frac{G^2_\text{F} V^2_{\text{ud}} (g^*_A)^2}{2\pi^3\hbar^7c^6} \mathcal{B}^{(\mp)}_i (-E_\nu - E^{(\mp)}_i)[(-E_\nu - E^{(\mp)}_i)^2-m_e^2c^4]^{1/2}E^2_\nu
  \\
\times [1-f_{e^{\mp}}(-E_\nu -E^{(\mp)}_i)] F(\pm Z+1, -E_\nu-E^{(\mp)}_i)\Theta(-E_\nu -E^{(\mp)}_i-m_ec^2),
\end{multline}
where upper (lower) sign corresponds to $\beta^-$- ($\beta^+$-)decay;
\item $\nu\bar\nu$-pair emission produces the same spectra for $\nu_e$  and $\bar\nu_e$ (The spectrum of other (anti)neutrino flavors is also given by~\eqref{spectrum_pair}.)
\begin{eqnarray}\label{spectrum_pair}
  \lambda_i^{\nu\bar\nu} (E_\nu)&=& \frac{G^2_\text{F}  g_A^2}{2\pi^3\hbar^7c^6} \mathcal{B}^{(0)}_i (-E_\nu - E^{(0)}_i)^2E^2_\nu \Theta(-E_\nu -E^{(0)}_i).
\end{eqnarray}
\end{itemize}

In the above expressions,  $G_\text{F}$ denotes the Fermi coupling constant,  $ V_{\text{ud}}$ is the up--down element of the Cabibbo--Kobayashi--Maskava quark-mixing matrix and $g_A=-1.27$ is the weak axial coupling constant. Note that, for charged current reactions, we use the effective coupling constant $g^*_A=0.74g_A$ that takes into account the observed quenching of the GT$_\pm$ strength. The~function  $f_{e^-(e^+)}(E)$ is  the Fermi--Dirac distribution for electrons (positrons), and~the Fermi function $F(Z,E)$ takes the distortion of the charged lepton wave function by the Coulomb field of the nucleus into account. It follows from the energy conservation that, for capture reactions, the electron (positron) energy is given by $E_{e^{\mp}} = E_\nu + E_i^{(\pm)}$,  while for the $\beta^{\pm}$-decay, we have $E_{e^{\mp}} + E_\nu = - E_i^{(\mp)}$, and~$E_{\bar\nu} + E_\nu = - E_i^{(0)}$ for $\nu\bar\nu$-pair emission. Obviously, only negative-energy transitions ($E_i^{(\pm,0)}<0$) contribute to $\beta^{\mp}$-decay and $\nu\bar\nu$-pair~emission.

Summation over different contributions $\mathrm{x} = \mathrm{EC,}\,\,\beta^+,\,\nu\bar\nu\,\,  (\mathrm{PC},\,\,\beta^-,\,\nu\bar\nu)$ and final states $i$ of a hot nucleus gives us the total (anti)neutrino spectrum
\begin{equation}\label{tot_spectr}
  \lambda(E_\nu) = \sum_{\mathrm{x}} \sum_i \lambda^{\mathrm{x}}_i (E_\nu).
\end{equation}
Then, the integration over $E_\nu$ yields the neutrino  emission ($\Lambda$) and energy-loss ($P$) rates
\begin{equation}
  \Lambda = \int \lambda(E_\nu)\,dE_\nu,~~~~P = \int \lambda(E_\nu) E_\nu dE_\nu.
\end{equation}

\section{Results}
\unskip

\subsection{Pre-Supernova~Model}

To study (anti)neutrino production and energy loss rates due to hot $^{56}$Fe in the pre-supernova environment, we use the model \texttt{25\_79\_0p005\_ml} from Farmer~et~al.~\cite{Farmer_ApJS227}. It is a typical
pre-supernova model with a good mass resolution and a core temperature that is
high enough for our estimates. Its name means that the initial mass of the model was
$25 M_\odot$, the~nuclear reaction network was \texttt{mesa\_79.net}, the~maximum mass of a
computational cell was $0.005 M_\odot$, and~the mass loss during the stellar
evolution was taken into account (see details in~\cite{Farmer_ApJS227}).

\begin{figure}[H]
\includegraphics[width=0.9\textwidth]{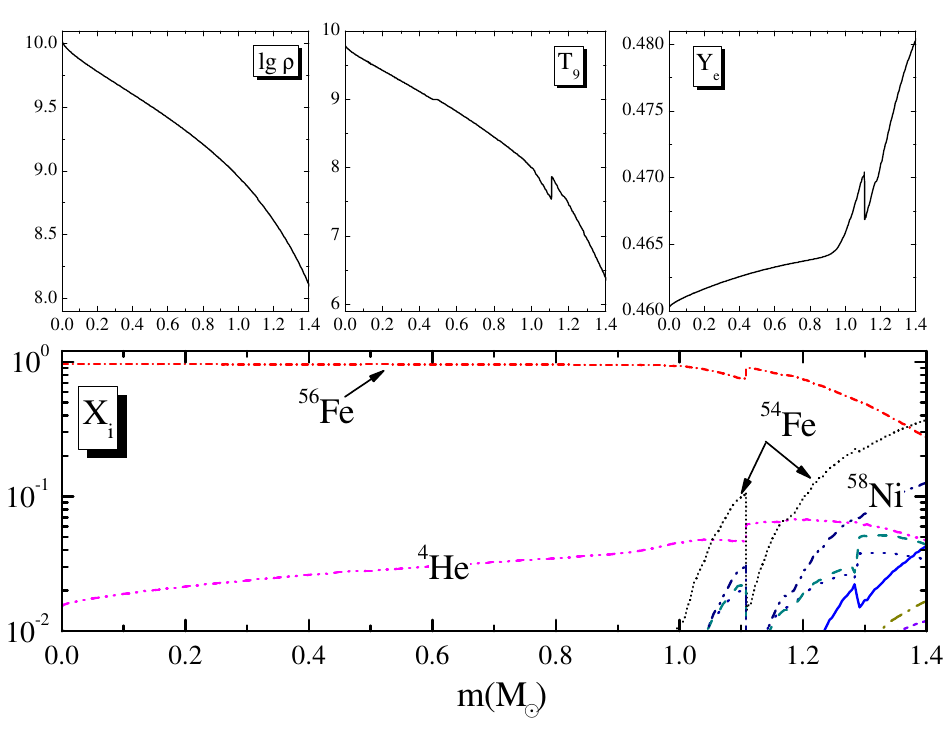}
\caption{\textbf{Top}  panels: density, temperature and electron fraction profiles along the mass coordinate for the  \texttt{25\_79\_0p005\_ml} pre-supernova model at the onset of the core collapse. \textbf{Bottom} panel: the respective mass fraction distribution of the most dominant~isotopes. \label{profile}}
\end{figure}

The authors of~\cite{Farmer_ApJS227} employed the stellar evolution code MESA~\cite{Paxton_ApJS220}, version 7624. In~output, MESA gives the time-evolving profiles of density $\rho$ (in g/ccm), temperature $T_9\equiv T(K)/10^9$, electron fraction $Y_\mathrm{e}$ and mass fraction $X_i$ of  various isotopes.
The profile that we use corresponds to the onset of core collapse, which is defined as the time when
the infall velocity exceeds 1000 km/s anywhere in the star.
The respective density, temperature and electron fraction profiles along the mass coordinate are demonstrated in the top panels of Figure~\ref{profile}.  In~the bottom panel of Figure~\ref{profile}, we show the mass fraction profiles of the most dominant isotopes. We  see that the $^{56}$Fe isotope is dominant up to $m<1.3M_\odot$. It is in this hot and dense central part of the star that the main neutrino flux is~born.

We calculate
(anti)neutrino spectra and energy loss rates due to hot $^{56}$Fe at six specific points on the mass coordinate for which $X_{^{56}\mathrm{Fe}}>0.5$.  To~select these points,  in~the MESA output file, we first identify the mass coordinate $m_{(6)}$ where $X_{^{56}\mathrm{Fe}}$ takes the value closest to 0.5. Then, the~remaining  five  points are taken from  the MESA output  and are uniformly distributed  over the interval $[0,m_{(6)}]$. The~values of $m_{(n)}$ with the respective values of the radial coordinate,  $^{56}$Fe mass fraction, density, temperature, electron fraction and electron  chemical potential   are given in Table~\ref{rhoTY}. It is clearly seen from  Table~\ref{rhoTY} that the range of temperature and density varies widely, while the electron fraction remains almost unchanged. The~resulting chemical potential reduces   four times when we move along the mass coordinate from point $m_{(1)}$ to $m_{(6)}$. Thus, the~selected  points  enable us to consider weak  nuclear processes under rather different representative pre-supernova~conditions.

\begin{table}[H]
\caption{Six specific points on the mass coordinate where the (anti)neutrino spectra and energy loss rates due to hot $^{56}$Fe are~computed.\label{rhoTY}}
\newcolumntype{C}{>{\centering\arraybackslash}X}
\begin{tabularx}{\textwidth}{cCCCCCCC}
\toprule
  \textbf{(n) } 	& \boldmath{$m\,(M_\odot)$}	& \boldmath{$R\,(R_\odot)$} & \boldmath{$X_{^{56}\mathrm{Fe}}$} & \boldmath{$T_9$}  & \boldmath{$\log(\rho)$} & \boldmath{$Y_\mathrm{e}$} & \boldmath{$\mu_\mathrm{e}$}\,\textbf{(MeV)~\textsuperscript{1}} \\
\midrule
(1)                  	& 1.953~$\times$~10$^{-6}$
 & 6.41~$\times$~10$^{-6}$ & 0.95822 & 9.79138  & 10.01954 &  0.46029 & 8.451\\
(2)	                    & 0.26005  & 3.75~$\times$~10$^{-4}$ & 0.95791 & 9.33784  &  9.72765 &	0.46197 & 6.679\\
(3)                   	& 0.51745  & 5.22~$\times$~10$^{-4}$ & 0.95872 & 8.95570  &  9.49678 &	0.46304 & 5.527\\
(4)	                    & 0.77778  & 6.71~$\times$~10$^{-4}$ & 0.95356 & 8.49247  &  9.23424 &	0.46379 & 4.435\\
(5)                  	& 1.03597  & 8.51~$\times$~10$^{-4}$ & 0.88566 & 7.86298  &  8.90200 &	0.46725 & 3.340\\
(6)                  	& 1.29568  & 1.13~$\times$~10$^{-3}$ & 0.497   & 6.97018  &  8.39942 &	0.47616 & 2.126\\
\bottomrule
\end{tabularx}
\noindent{\footnotesize{\textsuperscript{1} The chemical potential $\mu_\mathrm{e}$ is defined to include the rest mass so that $\mu_{\mathrm{e}^-}=-\mu_{\mathrm{e}^+}$.} The value of $\mu_\mathrm{e}$ is determined from the density $\rho Y_\mathrm{e}$ by inverting the relation $\rho Y_\mathrm{e} = (\pi^2\hbar^3 N_\mathrm{A})^{-1}\int\limits_0^\infty (f_{\mathrm{e}^-} {-} f_{\mathrm{e}^+})p^2 dp$.}
\end{table}
\vspace{-12pt}

\subsection{Thermal Effects on Gamow--Teller Strength Functions in $^{56}$Fe}

\begin{figure}[H]
\includegraphics[width=\textwidth]{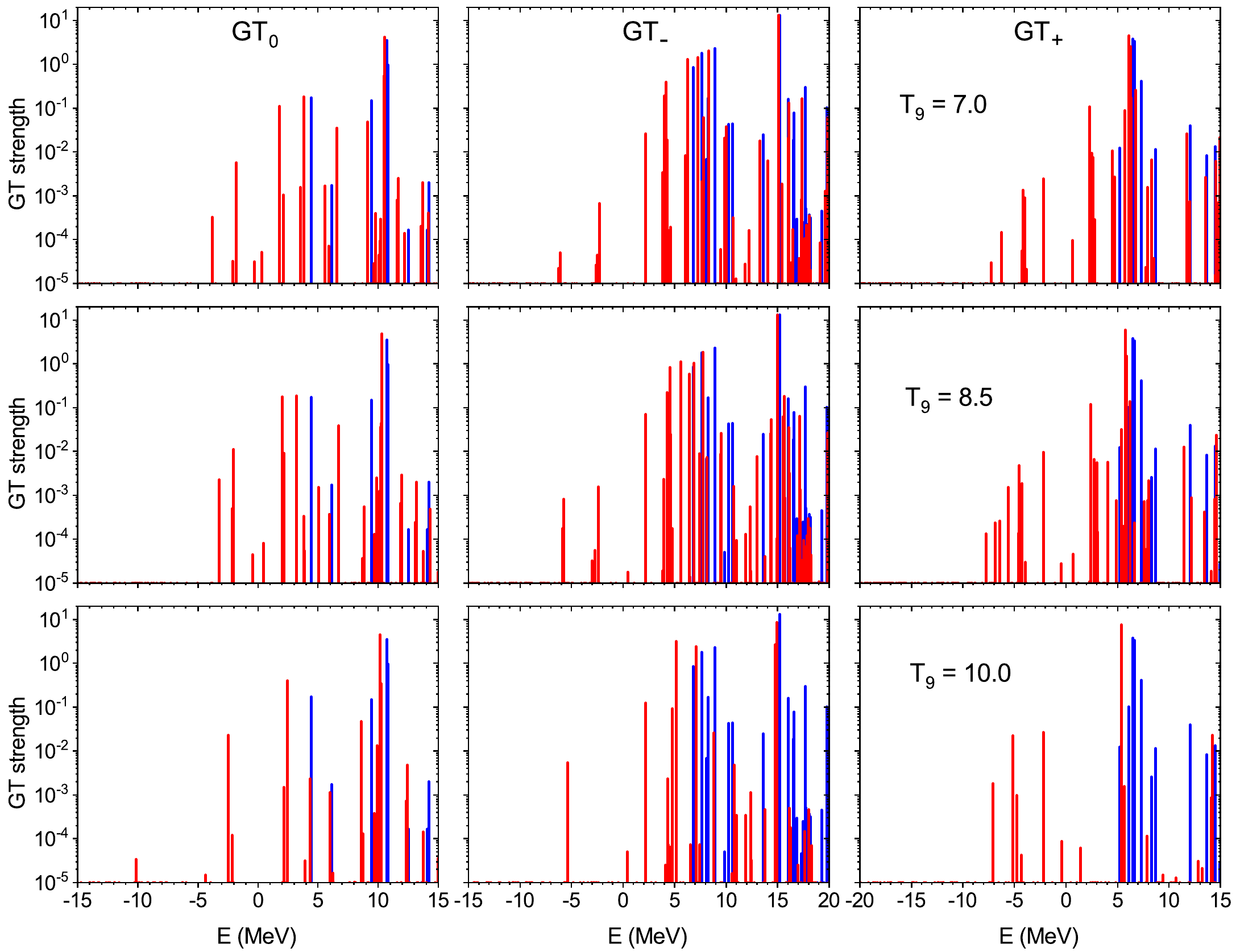}
\caption{GT$_0$ (\textbf{left} column), GT$_-$ (\textbf{middle}  column) and GT$_+$ (\textbf{right} column) strength functions $S_\mathrm{GT}$ for $^{56}$Fe calculated at $T_9=7.0$ (\textbf{upper} row), $T_9=8.5$ (\textbf{middle} row) and $T_9=10.0$ (\textbf{lower} row). The~blue bars represent the ground-state strength~functions.\label{GT_strength}}
\end{figure}

Before discussing (anti)neutrino spectra and energy loss rates,  we consider the thermal evolution  of the GT strength functions in $^{56}$Fe.
 In  Figure~\ref{GT_strength}, the~GT$_{0,\mp}$ strength functions  are displayed at three  temperatures relevant in the pre-supernova context. To~emphasize thermal effects, the~ground-state strength functions are also shown in each panel. The~choice of the nuclear model and its parameters  for TQRPA calculations is discussed in~\cite{Dzhioev_PhPN53_2}. Here, we just mention that the strength functions in Figure~\ref{GT_strength} are obtained by applying self-consistent calculations based on the SkM* parametrization of the Skyrme effective force. As~shown in~\cite{Dzhioev_PhPN53_2}, zero-temperature QRPA calculations with the  SkM* force fairly accurately reproduce both experimental data and shell-model results on the GT$_{0,\mp}$ resonance in  the ground state of $^{56}$Fe (TQRPA calculations performed with Skyrme forces SLy4, SkO' and SGII~\cite{Dzhioev_PRC81,Dzhioev_PRC92,Dzhioev_PRC101,Dzhioev_PRC89,Dzhioev_PRC94,Dzhioev_PhPN53_1,Dzhioev_PhPN53_2,Dzhioev_PhPN53_3} clearly demonstrate that thermal effects on the GT strength functions do not depend on a particular choice of the parametrization---for~this reason, all the results presented below  concerning the temperature dependence of (anti)neutrino spectra and energy loss rates  remain valid for other Skyrme parametrizations.). According to the present calculations, the~main contribution to the GT$_0$ resonance  ($E\approx15$\,MeV) in $^{56}$Fe comes from proton and neutron charge-neutral single-particle transitions $1f_{7/2}\to 1f_{5/2}$, while  the  GT$_-$ and GT$_+$ resonances at energies  of $E\approx15$\,MeV and  $E\approx 6$\,MeV, respectively,  are mainly formed by the $1f_{7/2}\to 1f_{5/2}$  charge-exchange~transitions.

As seen from the plots, the thermal effects can noticeably change the strength functions.  First, since our TQRPA calculations do not support  the Brink hypothesis,  the~GT strength for upward $(E>0)$  transitions exhibits a temperature dependence. Namely,  due to the vanishing of pairing correlations and thermal weakening of the residual  particle--hole
interaction,  the~GT$_{0,\mp}$ resonance   moves to lower energies.  Moreover, the~thermal smearing of the nuclear Fermi surfaces unblocks the low-energy GT transitions. In~charge-exchange strength functions $S_{GT_-}$ and $S_{GT_+}$, these transitions lead to the appearance  of the GT strength below the ground-state reaction threshold $Q$~(for $^{56}\text{Fe}\to{}^{56}\text{Mn}$ reactions $Q=4.207$\,MeV, and~for $^{56}\text{Fe}\to{}^{56}\text{Co}$  reactions $Q=4.055$\,MeV)
, while in the GT$_0$ distribution, finite temperature unblocks a low-energy strength below the experimental energy of the first $1^+$ state in $^{56}$Fe ($E_{1^+_1}\approx 3.12$\,MeV). Second,  a~temperature rise increases the population of nuclear excited
states and enables downward ($E<0$) transitions  in accordance with the detailed balance relations~(\ref{DB1}) and ( \ref{DB2}).
Comparing the GT$_-$ and GT$_+$ distributions at $T\ne0$, we see that the main contribution to the  negative-energy  GT$_-$ strength comes from the transition, which is inverse to the GT$_+$ resonance. At~the same time, the~main contribution to the GT$_+$ strength at $E<0$ comes from transitions inverse to low-energy GT$_-$ transitions, while  the contribution of the transition inverse to the GT$_-$ resonance is small. The~reason for this is that the GT$_-$ resonance is much higher in energy than the GT$_+$
resonance, and~therefore its  inverse  transition is strongly suppressed by the
Boltzmann exponential factor in the detailed balance relation~\eqref{DB2}. In~the GT$_0$ distribution, negative-energy transitions inverse to low-energy ones and to the excitation of the GT$_0$ resonance contribute to the downward~strength.

It is important to emphasize  that, since upward and downward strengths are connected by the detailed balance relation,  thermal effects on the upward GT strength also influence the  downward strength. In~\cite{Dzhioev_PRC89}, this influence was studied by comparing the  running (cumulative) sums for the GT$_0$ downward strength calculated using and without using the Brink hypothesis. In~particular, it was shown that both the thermal unblocking of low-energy strength and lowering the GT resonance significantly enhance the strength of negative-energy transitions. Eventually, this enhancement should have important consequences for (anti)neutrinos emitted due  de-excitation and decay~processes.

\subsection{(Anti)neutrino Spectra and Energy Loss Rates}

\begin{figure}[H]
\includegraphics[width=\textwidth]{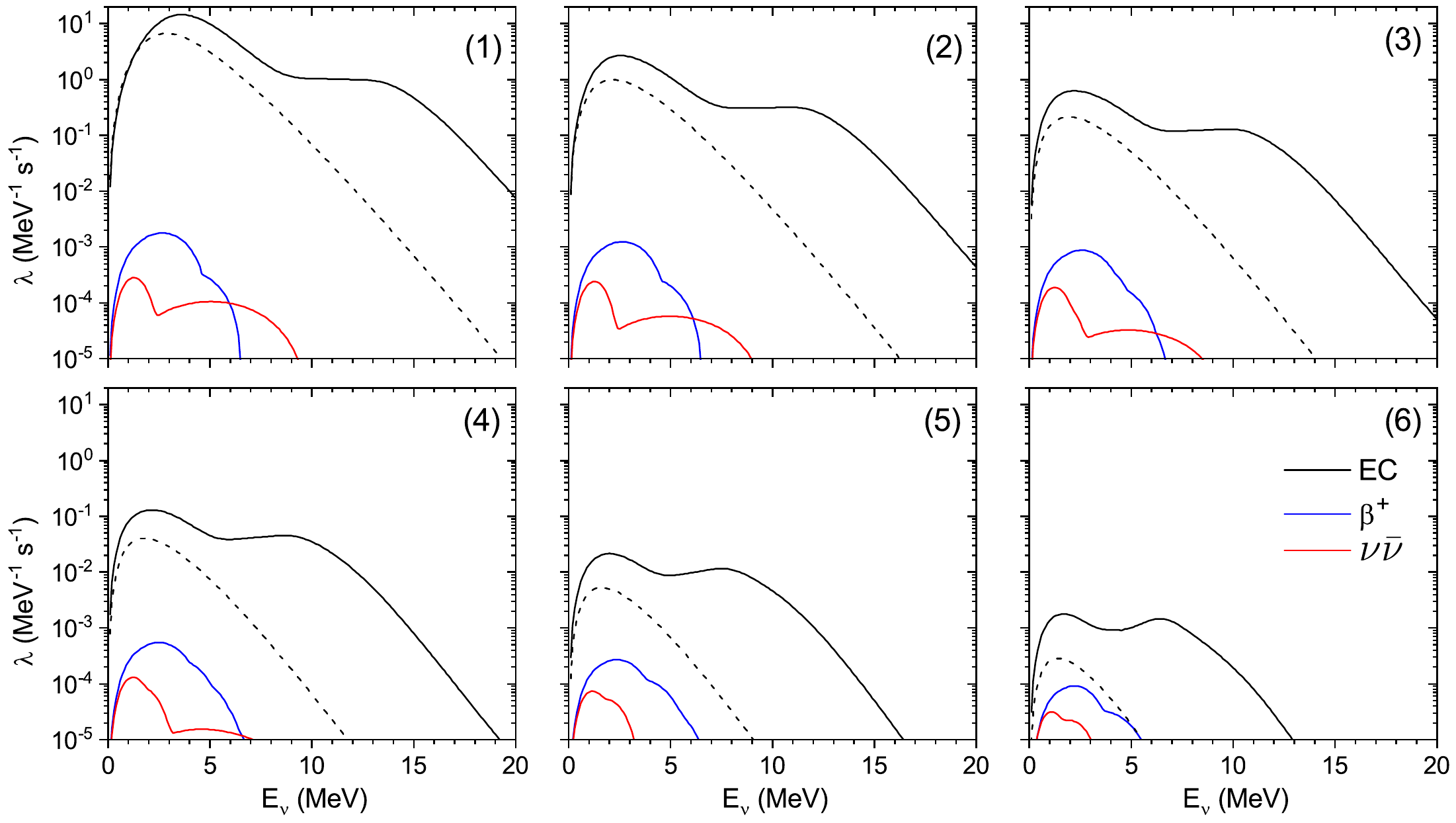}
\caption{Neutrino spectra produced by $^{56}$Fe due to electron capture (EC), $\beta^+$-decay  and $\nu\bar\nu$-pair emission. Each set of curves corresponds to a specific point (n) (n=1,\,2,\,3,\,4,\,5,\,6) on the mass coordinate  listed in Table~\ref{rhoTY}. The~dashed curves represent neutrino spectra arising from EC on the ground state of  $^{56}$Fe. \label{neutrino}}
\end{figure}

We now demonstrate the $\nu_e$ and $\bar\nu_e$ energy spectra for six selected points inside the star. Figure~\ref{neutrino} shows the contribution of different nuclear weak processes to  neutrino spectra for each  point from Table~\ref{rhoTY}. Although~the shape and intensity of the spectrum depend on the temperature, density and electron fraction, there are features common for all points. Namely, for~all  points, the spectra are dominated by the EC contribution that exhibits a low-energy peak and a high-energy tail. The~latter gradually transforms into the second peak when we move from the center of the star. Our analysis shows that low-energy neutrinos are emitted after electron capture excites  the GT$_+$ resonance state, while high-energy neutrinos are caused by thermally unblocked low- and negative-energy GT$_+$ transitions.

In Figure~\ref{neutrino}, the~importance of thermal effects is illustrated by comparing neutrino spectra produced by hot $^{56}$Fe with that produced  by a cold nucleus, when only EC is possible.  As~clearly seen from  Figure~\ref{neutrino},  the thermal unblocking of the GT$_+$ strength  at $E<0$ (see the right panels in Figure~\ref{GT_strength}) leads to the appearance of  high-energy neutrinos  in the spectra, whose fraction increases when we move from point (1) to (6). Moreover, as~shown in the figure, the temperature-induced lowering of the GT$_+$ resonance amplifies the low-energy ($E_{\nu_e}<5$\,MeV) part of the spectra  and shifts its maximum to higher~energies.

\begin{figure}[H]
\includegraphics[width=\textwidth]{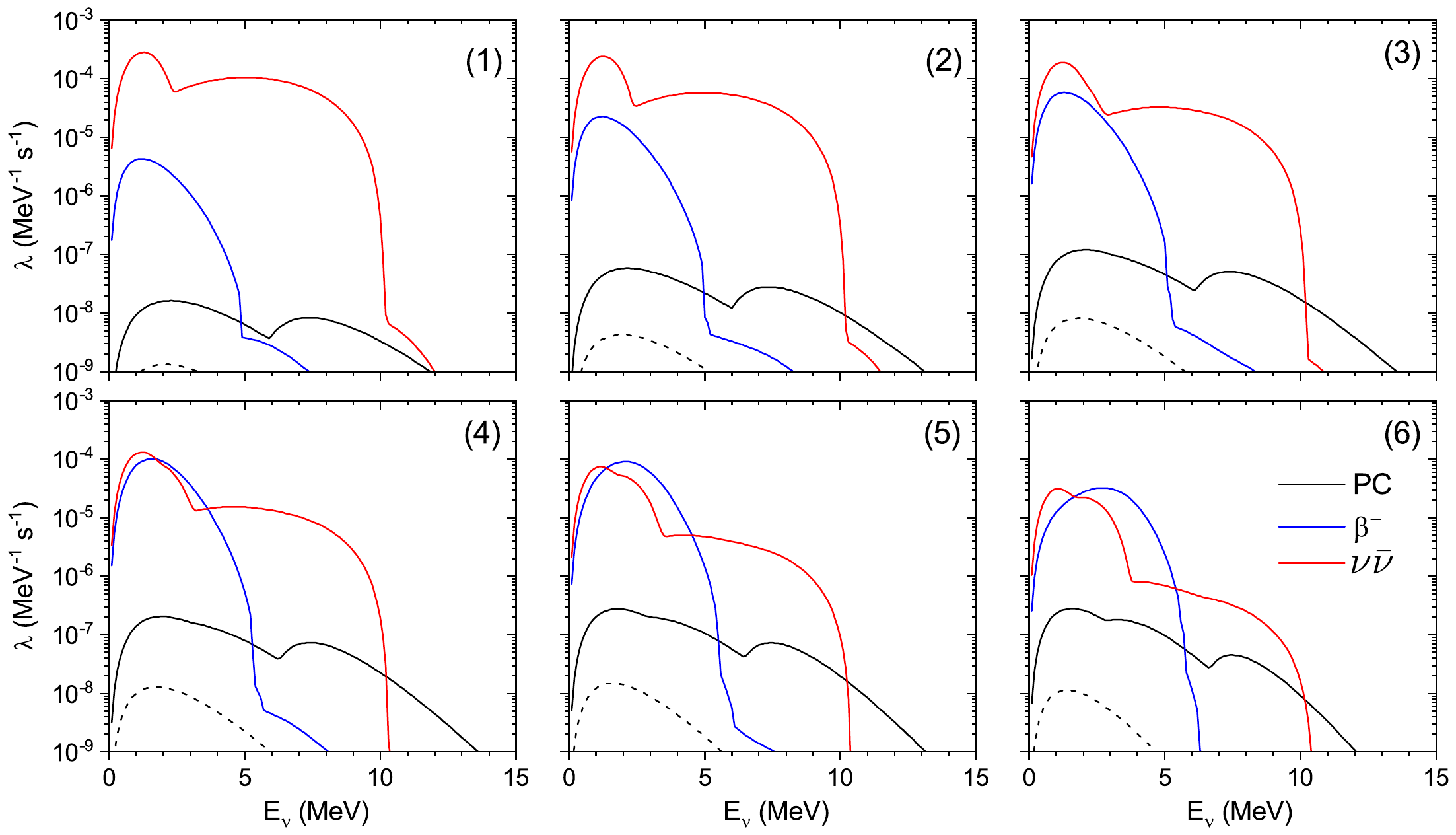}
\caption{Antineutrino  spectra produced by $^{56}$Fe due to positron capture (PC), $\beta^-$-decay and $\nu\bar\nu$-pair emission. Each set of curves corresponds to a specific point (n) (n=1,\,2,\,3,\,4,\,5,\,6) on the mass coordinate  listed in Table~\ref{rhoTY}. The~dashed curves represent antineutrino spectra arising from PC on the ground state of  $^{56}$Fe. \label{antineutrino}}
\end{figure}

The contribution of different nuclear weak processes to the antineutrino spectra produced by hot $^{56}$Fe is shown in Figure~\ref{antineutrino}. Our calculations clearly demonstrate the dominance of $\nu_e\bar\nu_e$-pair emission in the antineutrino spectra under pre-supernova  conditions when the $\beta^-$-decay is strongly blocked by the electron chemical potential. The~obtained $\nu_e\bar\nu_e$-spectra have a narrow low-energy peak at $E_\nu\approx 1-2$\,MeV and a broad high-energy one peaking around $E_\nu\approx 5$\,MeV. By~matching with the GT$_0$ strength function in Figure~\ref{GT_strength}, we conclude that the former arises due to thermally unblocked low-energy downward GT$_0$ transitions, while high-energy antineutrinos are emitted from the $\nu_e\bar\nu_e$-decay of the thermally populated GT$_0$ resonance. Since the  GT$_0$ resonance in $^{56}$Fe is located at relatively high energy, its thermal population rapidly decreases at low temperatures, leading to a decrease in the fraction of high-energy antineutrinos. Nevertheless, amongst weak nuclear processes, it is the $\nu_e\bar\nu_e$-decay of the   GT$_0$ resonance that produces the high-energy antineutrinos of all flavors under pre-supernova conditions listed in~Table~\ref{rhoTY}. It is also seen from Figure~\ref{antineutrino} that temperature reduction  has a modest impact on the intensity of low-energy antineutrinos emitted due to the $\nu_e\bar\nu_e$-decay. At~the same time, the~reduction in the chemical potential $\mu_e$ unblocks $\beta^-$-decay, which  also emits low-energy  antineutrinos. As~a result, when we move from the center of the star, the contributions of the $\nu_e\bar\nu_e$-pair emission and $\beta^-$-decay to the low-energy antineutrino spectrum become~comparable.

\begin{figure}[H]
\includegraphics[width=0.9\textwidth]{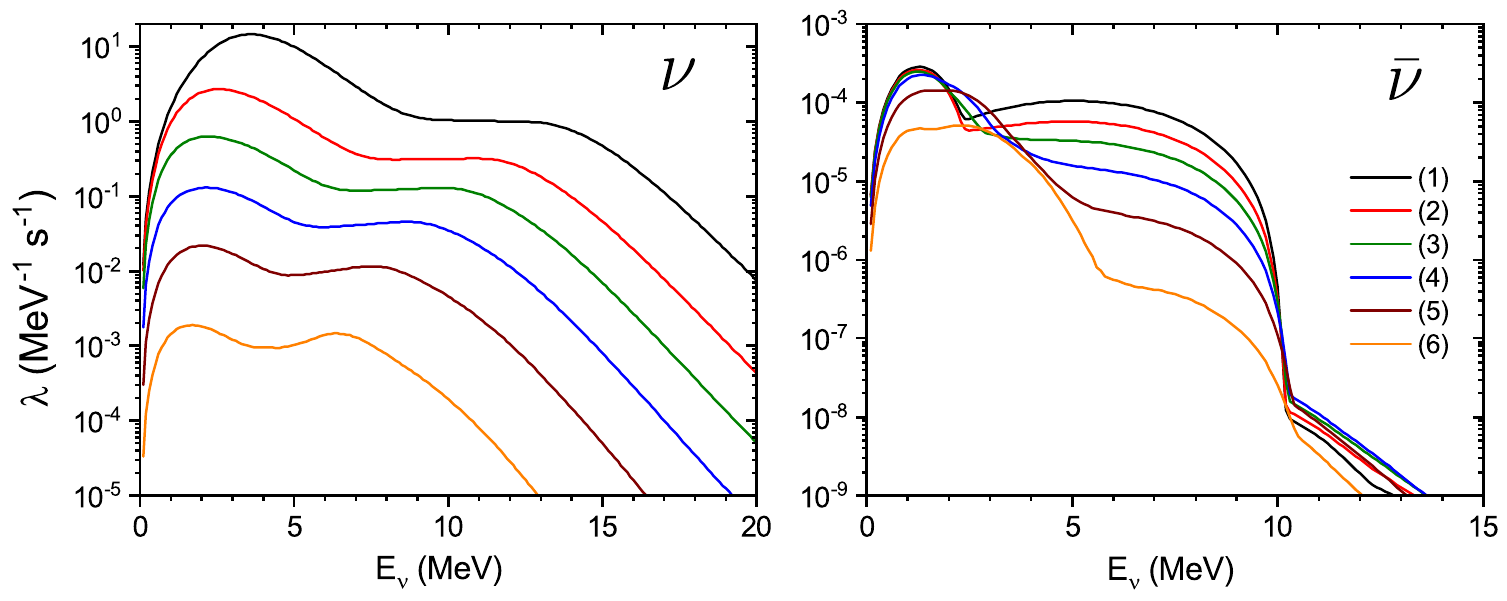}
\caption{Neutrino and antineutrino spectra $\lambda$ due to weak processes with hot $^{56}$Fe for specific points (n) (n=1,\,2,\,3,\,4,\,5,\,6) on the mass coordinate listed in Table~\ref{rhoTY}.
   \label{spectrum}}
\end{figure}

Figure~\ref{spectrum} shows the evolution of the total (anti)neutrino spectrum $\lambda$~\eqref{tot_spectr} as we move from the center of the star.
Since electron capture is a dominant source of neutrinos, the~reduction in the chemical potential $\mu_\mathrm{e}$ below the GT$_+$ resonance  energy decreases the low-energy  peaks in $\lambda$ by about four orders of magnitude, while the high-energy tail is reduced by approximately three orders of magnitude. For~this reason, a relative fraction of high-energy neutrinos in the spectrum increases.   As~discussed  above, contributions from the $\nu\bar\nu$-pair emission and $\beta^-$-decay to emission of low-energy antineutrinos  demonstrate opposite trends when we move from points (1) to (6). Therefore, the~intensity of the low-energy antineutrino emission is rather unsensitive to the change in pre-supernova conditions.  At~the same time, the~intensity of high-energy antineutrino emission is reduced by more than two orders of magnitude as the temperature decreases from $T_9\approx 9.8$ to $T_9\approx 7.0$.

\begin{figure}[H]
\includegraphics[width=\textwidth]{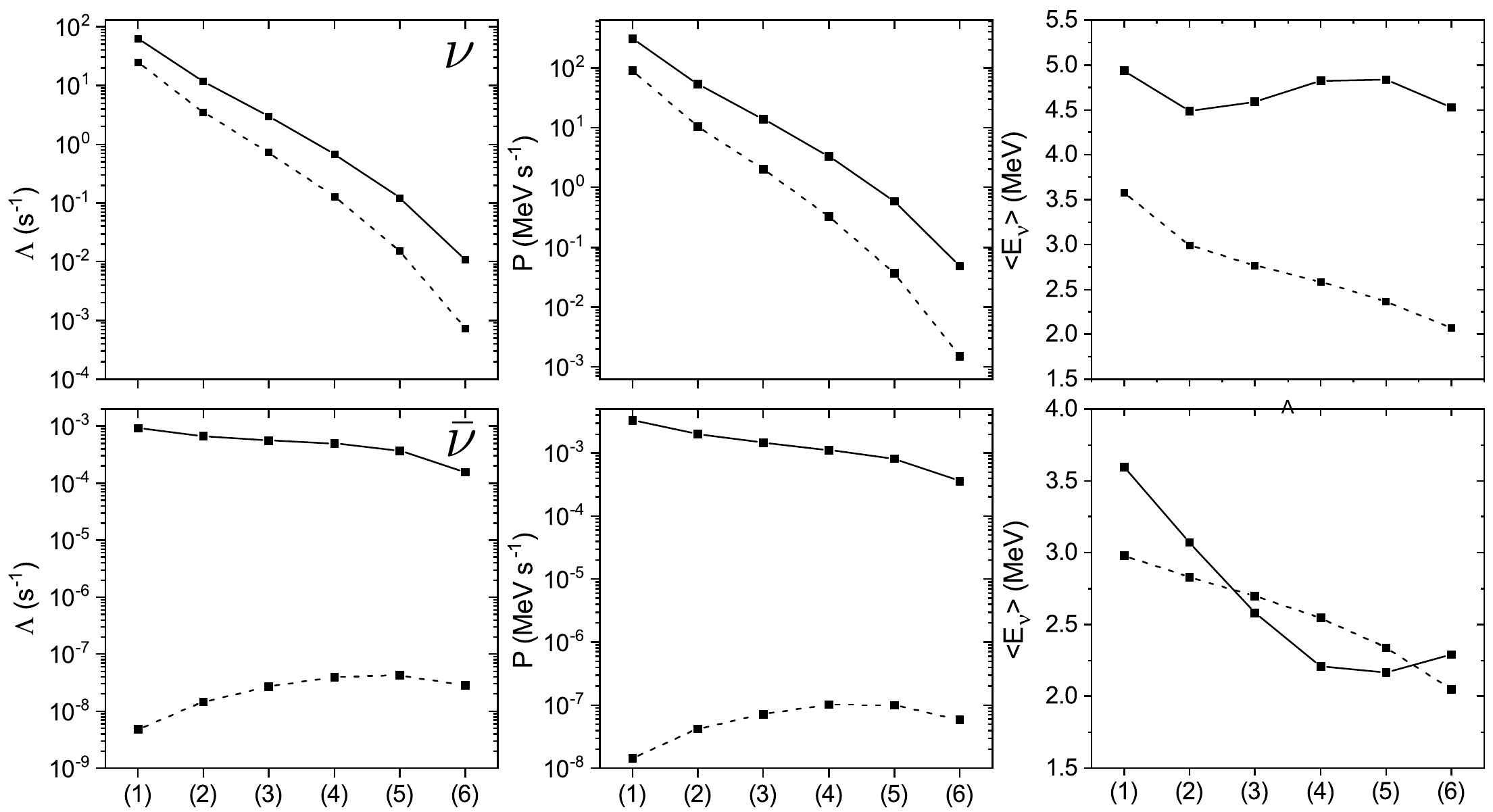}
\caption{Neutrino (\textbf{top}) and antineutrino (\textbf{bottom}) emission rate $\Lambda$, energy-loss rate $P$ and~average energy $\langle E_\nu\rangle$ due to hot $^{56}$Fe  for specific points (n) (n=1,\,2,\,3,\,4,\,5,\,6) on the mass coordinate listed in Table~\ref{rhoTY}.  The~dashed curves show $\Lambda$, $P$,  and~$\langle E_\nu\rangle$  calculated for the ground state of  $^{56}$Fe.  \label{total}}
\end{figure}

In Figure~\ref{total},   the~emission rates $\Lambda$, energy-loss rates $P$, and~the average energy $\langle E_\nu\rangle = P/\Lambda$ for the electron (anti)neutrinos emitted due to weak processes with hot $^{56}$Fe are shown. Referring to the figure, neutrino rates demonstrate a strong dependence under pre-supernova conditions. Compared with the ground-state rates,
we conclude that temperature lowering gives a minor contribution to a severe reduction in the neutrino rates and the latter is mainly caused by the chemical potential decrease.
In contrast, as~pair emission  only  depends on temperature and $\beta^-$-decay rate increases when $\mu_\mathrm{e}$ decreases,  the~computed antineutrino rates demonstrate a more modest  dependence  under pre-supernova conditions. We also see that the finite temperature of the nucleus plays a more important role for antineutrino rates than for neutrino~ones.

\unskip

As for the average energy, for~emitted neutrinos, it varies rather weakly around $\langle E_\nu\rangle\approx 4.7$\,MeV. This stability is a result of the increasing fraction of high-energy neutrinos emitted by de-excitation processes, which compensates the decrease in available electron energy when we move from the center of the star. This is clearly seen if we compute $\langle E_\nu\rangle$ for the cold $^{56}$Fe. In~that case, $\langle E_\nu\rangle$ is essentially lower and shows a decreasing trend. At~the same time, the~average energy of antineutrinos demonstrates  non-monotonic behavior due to the competition between $\nu\bar\nu$-decay and $\beta^-$-decay. Moreover, since, in decay processes, the released energy is shared among two emitted particles, the~antineutrino average energy is smaller than that for~neutrinos.

\section{Discussion and~Perspectives}

Neutrino spectra shown in Figure~\ref{neutrino} confirm the conclusion of Ref.~\cite{Langanke_PRC64} that the single-strength approximation can be applied under stellar conditions  with the electron chemical potential high enough to allow  the excitation of the GT$_+$ resonance by electron capture. Such conditions occur during the collapse phase. However, our calculations clearly demonstrate that this approximation can fail in the pre-supernova phase when negative-energy GT$_+$ transitions from thermally excited states noticeably contribute to electron capture and the resulting neutrino energy spectrum is double-peaked. On~the whole, the~present thermodynamically consistent calculations of electron neutrino spectra  performed without assuming the Brink hypothesis  indicate that the thermal effects on the GT$_+$ strength function  shift the spectrum to higher energies, and~thus make the neutrino detection more~likely.

The inclusion of $\nu\bar\nu$-pair emission into consideration shows that this neutral current process might be a dominant source of high-energy antineutrinos emitted via the de-excitation of the GT$_0$ resonance. Considering that the energy of the GT$_0$ resonance is related to the spin-orbit splitting, the~high-energy peak in antineutrino spectra can be easily parameterized. Moreover, since the $\nu\bar\nu$-pair emission only depends on temperature, the~detection of high-energy pre-supernova antineutrinos might be a test for thermodynamic conditions in the stellar~interior.

The next evident step in our study of the role of nuclear weak processes in pre-supernova (anti)neutrino production is to compute overall (anti)neutrino spectra and energy loss rates as well as their time evolution for different stellar progenitors. To~this end, calculations such as those performed for $^{56}$Fe are needed for isotopes abundant in the  stellar core and then, in the integration over the whole core, these should be performed for several time steps. Concerning the possibility of (anti)neutrino detection, we should take into account (ant)neutrino flavor oscillation, which changes the initial flavor composition of the pre-supernova (anti)neutrino~flux.

\authorcontributions{Conceptualization: A.A.D. and A.V.Y.;
                      formal analysis:  A.A.D., A.V.Y., N.V.D.-B. and A.I.V.;
                      software: A.A.D., A.V.Y. and N.V.D.-B.;
                      writing---original draft preparation: A.A.D.;
                      writing---review and editing: A.A.D., A.I.V., A.V.Y. and N.V.D.-B.
                      All authors have read and agreed to the published version of the~manuscript.}

\funding{A.V.Y. thanks RSF 21-12-00061 grant for~support.}

\dataavailability{Not applicable.}

\conflictsofinterest{The authors declare no conflict of~interest.}

\begin{adjustwidth}{-\extralength}{0cm}

\reftitle{References}



\begin{thebibliography}{999}

\bibitem[Woosley et~al.(2002)Woosley, Heger, and Weaver]{Woosley_RMPh74}
Woosley, S.E.; Heger, A.; Weaver, T.A.
\newblock {The evolution and explosion of massive stars}.
\newblock {\em Rev. Mod. Phys.} {\bf 2002}, {\em 74},~1015--1071.
\newblock {\url{https://doi.org/10.1103/RevModPhys.74.1015}}.

\bibitem[Janka et~al.(2007)Janka, Langanke, Marek, Mart\'{\i}nez-Pinedo, and
  M{\"{u}}ller]{Janka_PhysRep442}
Janka, H.T.; Langanke, K.; Marek, A.; Mart\'{\i}nez-Pinedo, G.; M{\"{u}}ller,
  B.
\newblock {Theory of core-collapse supernovae}.
\newblock {\em Phys. Rep.} {\bf 2007}, {\em 442},~38--74.
\newblock
  {\url{https://doi.org/http://dx.doi.org/10.1016/j.physrep.2007.02.002}}.

\bibitem[Balasi et~al.(2015)Balasi, Langanke, and
  Mart\'{\i}nez-Pinedo]{Balasi_PPNP85}
Balasi, K.G.; Langanke, K.; Mart\'{\i}nez-Pinedo, G.
\newblock {Neutrino-nucleus reactions and their role for supernova dynamics and
  nucleosynthesis}.
\newblock {\em Prog. Part. Nucl. Phys.} {\bf 2015}, {\em
  85},~33--81.
\newblock {\url{https://doi.org/http://dx.doi.org/10.1016/j.ppnp.2015.08.001}}.

\bibitem[Kato et~al.(2020)Kato, Ishidoshiro, and Yoshida]{Kato_ARNPS70}
Kato, C.; Ishidoshiro, K.; Yoshida, T.
\newblock {Theoretical prediction of presupernova neutrinos and their
  detection}.
\newblock {\em Annu. Rev. Nucl. Part. Sci.} {\bf 2020}, {\em
  70},~121--145.  
\newblock {\url{https://doi.org/10.1146/annurev-nucl-040620-021320}}.

\bibitem[Patton et~al.(2017{\natexlab{a}})Patton, Lunardini, and
  Farmer]{Patton_ApJ840}
Patton, K.M.; Lunardini, C.; Farmer, R.J.
\newblock {Presupernova Neutrinos: Realistic Emissivities from Stellar
  Evolution}.
\newblock {\em  Astrophys. J.} {\bf 2017}, {\em 840},~2.
\newblock {\url{https://doi.org/10.3847/1538-4357/aa6ba8}}.

\bibitem[Patton et~al.(2017{\natexlab{b}})Patton, Lunardini, Farmer, and
  Timmes]{Patton_ApJ851}
Patton, K.M.; Lunardini, C.; Farmer, R.J.; Timmes, F.X.
\newblock {Neutrinos from Beta Processes in a Presupernova: Probing the
  Isotopic Evolution of a Massive Star}.
\newblock {\em  Astrophys. J.} {\bf 2017}, {\em 851},~6.
\newblock {\url{https://doi.org/10.3847/1538-4357/aa95c4}}.

\bibitem[Langanke et~al.(2001)Langanke, Mart{\'{i}}nez-Pinedo, and
  Sampaio]{Langanke_PRC64}
Langanke, K.; Mart{\'{i}}nez-Pinedo, G.; Sampaio, J.M.
\newblock {Neutrino spectra from stellar electron capture}.
\newblock {\em Phys. Rev. C-Nucl. Phys.} {\bf 2001}, {\em 64},~055801.
\newblock {\url{https://doi.org/10.1103/PhysRevC.64.055801}}.

\bibitem[Misch and Fuller(2016)]{Misch_PRC94}
Misch, G.W.; Fuller, G.M.
\newblock {Nuclear neutrino energy spectra in high temperature astrophysical
  environments}.
\newblock {\em Phys. Rev. C} {\bf 2016}, {\em 94},~55808.
\newblock {\url{https://doi.org/10.1103/PhysRevC.94.055808}}.

\bibitem[Dzhioev et~al.(2010)Dzhioev, Vdovin, Ponomarev, Wambach, Langanke, and
  Mart{\'{i}}nez-Pinedo]{Dzhioev_PRC81}
Dzhioev, A.A.; Vdovin, A.I.; Ponomarev, V.Y.; Wambach, J.; Langanke, K.;
  Mart{\'{i}}nez-Pinedo, G.
\newblock {Gamow-Teller strength distributions at finite temperatures and
  electron capture in stellar environments}.
\newblock {\em Phys. Rev. C} {\bf 2010}, {\em 81},~015804.
\newblock {\url{https://doi.org/10.1103/PhysRevC.81.015804}}.

\bibitem[Dzhioev et~al.(2015)Dzhioev, Vdovin, and Wambach]{Dzhioev_PRC92}
Dzhioev, A.A.; Vdovin, A.I.; Wambach, J.
\newblock {Neutrino absorption by hot nuclei in supernova environments}.
\newblock {\em Phys. Rev. C} {\bf 2015}, {\em 92},~045804.
\newblock {\url{https://doi.org/10.1103/PhysRevC.92.045804}}.

\bibitem[Dzhioev et~al.(2020)Dzhioev, Langanke, Mart{\'{i}}nez-Pinedo, Vdovin,
  and {Stoyanov Ch.}]{Dzhioev_PRC101}
Dzhioev, A.A.; Langanke, K.; Mart{\'{i}}nez-Pinedo, G.; Vdovin, A.I.; {Stoyanov
  Ch.}.
\newblock {Unblocking of stellar electron capture for neutron-rich $N=50$
  nuclei at finite temperature}.
\newblock {\em Phys. Rev. C} {\bf 2020}, {\em 101},~025805.
\newblock {\url{https://doi.org/10.1103/PhysRevC.101.025805}}.

\bibitem[Dzhioev et~al.(2014)Dzhioev, Vdovin, Wambach, and
  Ponomarev]{Dzhioev_PRC89}
Dzhioev, A.A.; Vdovin, A.I.; Wambach, J.; Ponomarev, V.Y.
\newblock {Inelastic neutrino scattering off hot nuclei in supernova
  environments}.
\newblock {\em Phys. Rev. C} {\bf 2014}, {\em 89},~035805.
\newblock {\url{https://doi.org/10.1103/PhysRevC.89.035805}}.

\bibitem[Dzhioev et~al.(2016)Dzhioev, Vdovin, Mart{\'{i}}nez-Pinedo, Wambach,
  and {Stoyanov Ch.}]{Dzhioev_PRC94}
Dzhioev, A.A.; Vdovin, A.I.; Mart{\'{i}}nez-Pinedo, G.; Wambach, J.; {Stoyanov
  Ch.}.
\newblock {Thermal quasiparticle random-phase approximation with Skyrme
  interactions and supernova neutral-current neutrino-nucleus reactions}.
\newblock {\em Phys. Rev. C} {\bf 2016}, {\em 94},~015805.
\newblock {\url{https://doi.org/10.1103/PhysRevC.94.015805}}.

\bibitem[Dzhioev and Vdovin(2022{\natexlab{a}})]{Dzhioev_PhPN53_1}
Dzhioev, A.A.; Vdovin, A.I.
\newblock {Superoperator Approach to the Theory of Hot Nuclei and Astrophysical
  Applications: I---Spectral Properties of Hot Nuclei}.
\newblock {\em Phys. Part. Nucl.} {\bf 2022}, {\em 53},~885--938.
\newblock {\url{https://doi.org/10.1134/S1063779622050033}}.

\bibitem[Dzhioev and Vdovin(2022{\natexlab{b}})]{Dzhioev_PhPN53_2}
Dzhioev, A.; Vdovin, A.I.
\newblock {Superoperator Approach to The Theory of Hot Nuclei and Astrophysical
  Applications: II -- Electron Capture in Stars}.
\newblock {\em Phys. Part. Nucl.} {\bf 2022}, {\em 53},~939--999.
\newblock {\url{https://doi.org/10.1134/S1063779622050045}}.

\bibitem[Dzhioev and Vdovin(2022{\natexlab{c}})]{Dzhioev_PhPN53_3}
Dzhioev, A.A.; Vdovin, A.I.
\newblock {Superoperator Approach to the Theory of Hot Nuclei and Astrophysical
  Applications. III: Neutrino–Nucleus Reactions in Stars}.
\newblock {\em Phys. Part. Nucl.} {\bf 2022}, {\em
  53},~1051--1088.
\newblock {\url{https://doi.org/10.1134/S106377962206003X}}.

\bibitem[Farmer et~al.(2016)Farmer, Fields, Petermann, Dessart, Cantiello,
  Paxton, and Timmes]{Farmer_ApJS227}
Farmer, R.; Fields, C.E.; Petermann, I.; Dessart, L.; Cantiello, M.; Paxton,
  B.; Timmes, F.X.
\newblock {On Variations of Pre-Supernova Model Properties}.
\newblock {\em  Astrophys. J. Suppl. Ser.} {\bf 2016}, {\em
  227},~22.
\newblock {\url{https://doi.org/10.3847/1538-4365/227/2/22}}.

\bibitem[Paxton et~al.(2015) Paxton, Marchant, Schwab, Bauer, Bildsten, Cantiello, Dessart, Farmer, Hu, Langer, Townsend, Townsley, Timmes]{Paxton_ApJS220}
 Paxton, B.; Marchant, P.; Schwab, J.; Bauer, E.B.; Bildsten, L.; Cantiello, M.; Dessart, L.; Farmer, R.; Hu, H.; Langer, N.; Townsend, R.H.D.; Townsley, D.M.; Timmes, F.X.
\newblock {Modules for Experiments in Stellar Astrophysics (MESA): Binaries, Pulsations, and Explosions}.
\newblock {\em  Astrophys. J. Suppl. Ser.} {\bf 2015}, {\em
  220},~15.
\newblock {\url{https://doi.org/10.1088/0067-0049/220/1/15}}.

\end{thebibliography}

\PublishersNote{}
\end{adjustwidth}
\end{document}